\documentclass[smallextended,nospthms,final]{svjour3}
\usepackage[utf8]{inputenc}
\usepackage[T1]{fontenc}
\usepackage{amsmath}
\usepackage{amssymb}
\usepackage{mathptmx}
\usepackage{graphicx}
\usepackage{subfigure}
\usepackage{mathrsfs}
\usepackage{microtype}
\usepackage{cite}

\numberwithin{equation}{section}

\graphicspath{{./ArticleSubmission/}}

\newcommand*{\scri}{\mathscr{I}}
\newcommand*{\dd}{\mathrm{d}}
\newcommand*{\eths}{\eth_0}

\newcommand*{\del}{\partial}

\journalname{GRG}

\title{The second order spin-2 system in flat space near space-like and null-infinity}
\author{Georgios Doulis \and J\"org Frauendiener}
\institute{Georgios Doulis \at Department of Mathematics and Statistics, University of Otago, Dunedin 9010, New Zealand\\\email{gdoulis@maths.otago.ac.nz}
\and
J\"org Frauendiener \at Department of Mathematics and Statistics, University of Otago, Dunedin 9010, New Zealand\\\email{joergf@maths.otago.ac.nz}
}
\date{\today}

\begin{document}

\maketitle

\begin{abstract}
  In previous work, the numerical solution of the linearized gravitational field equations
  near space-like and null-infinity was discussed in the form of the spin-2 zero-rest-mass
  equation for the perturbations of the conformal Weyl curvature. The motivation was to
  study the behavior of the field and properties of the numerical evolution of the system
  near infinity using Friedrich's conformal representation of space-like infinity as a
  cylinder. It has been pointed out by H.O. Kreiss and others that the numerical evolution
  of a system using second order wave equations has several advantages compared to a
  system of first order equations. Therefore, in the present paper we derive a system of second
  order wave equations and prove that the solution spaces of the two systems are the same
  if appropriate initial and boundary data are given. We study the properties of this
  system of coupled wave equations in the same geometric setting and discuss the
  differences between the two approaches.
\end{abstract}

\section{Introduction}
\label{sec:intro}

In a recent paper~\cite{Beyer:2012ti} we studied the behaviour of gravitational
perturbations on Minkowski space near space-like infinity. For this purpose we used the
spin-2 zero-rest-mass equation which governs the perturbations of the Weyl curvature. The
system was studied using a conformal gauge given by Friedrich~\cite{Friedrich:1998tc}
which represents space-like infinity as a cylinder connecting past and future
null-infinity. This gauge exhibits explicitly the main issues of the propagation of
gravitational waves in asymptotically flat space-times, namely the question of how
$\scri^+$ is related to $\scri^-$, or, in other words, how the late ingoing gravitational
radiation affects the early outgoing radiation.

In~\cite{Beyer:2012ti} it was shown how the field propagates near the cylinder. It turns
out that the cylinder is a total characteristic in the sense that the equations turn into
intrinsic equations on it, there appear no outward derivatives. Hence, the field on the
cylinder is entirely determined by initial data on a space-like initial hyper-surface and
there are no free data to be prescribed on the cylinder.

The crucial property, however, is the fact that the intrinsic equations degenerate on the
intersections $I^\pm$ between the cylinder and null-infinity $\scri^\pm$. Here, the field
may develop singularities which can propagate onto null-infinity destroying its
smoothness. Friedrich has conjectured that certain conditions on the initial data 
have to be satisfied in order for the field to be smooth on $\scri$.

These analytical properties have been explored numerically in~\cite{Beyer:2012ti} based on
the first order system for the spin-2 field. The behavior of the resulting numerical
solutions near the cylinder $I$ was studied.  The performance of the code near the
ill-behaved location $I^+$ at the junction of the cylinder and null-infinity was tested
using different initial data. The analytical results were reproduced quite
successfully. In particular, the analytically predicted singular behaviour could be
reproduced numerically. The results in \cite{Beyer:2012ti} strongly indicate that
appropriate classes of initial data can be successfully evolved without loss of
convergence up and including the region $I^+$, but not beyond. As expected, since the
equations are no longer hyperbolic beyond $I^+$ the code develops numerical instabilities
and crashes.

With the present article we want to follow up on this work using an equivalent system of
second order wave equations. Our main motivation for this is the claim by several
numerical analysts (see e.g., Kreiss and Ortiz~\cite{Kreiss:2002du}) that the numerical
solutions of second order wave equations have better properties than those for the
corresponding first order systems. Specifically, according to \cite{Kreiss:2002du}, numerical 
approximations based on second order equations lead to better accuracy than the ones based on 
first order equations. In addition, in the second order case, spurious high-frequency waves travelling 
against the characteristics disappear.

The structure of the paper is as follows. In sec.~\ref{sec:minkowski_near_i} we briefly
describe Friedrich's representation of space-like infinity of Minkowski space as a
cylinder. In sec.~\ref{sec:spin-2_wave_system} we derive the system of wave equations and
show under what conditions it is equivalent to the first order system used
in~\cite{Beyer:2012ti}. In sec.~\ref{sec:numerical_analysis} our numerical implementation
and results are presented.  Finally, we conclude with a discussion in
sec.~\ref{sec:disscusion}. The conventions used in this work are those
of~\cite{Penrose:1984wr}.

\section{The finite representation of Minkowski space-time near space-like infinity}
\label{sec:minkowski_near_i}

In this section, we will briefly summarize the finite representation of Minkowski space-time close to 
space-like infinity originally proposed by Friedrich in \cite{Friedrich:1998tc}; more detailed 
discussions of this topic can be found in \cite{Beyer:2012ti,Friedrich:1998tc,ValienteKroon:2002dx}.

\begin{figure}[htb]
  \centering
  \includegraphics[width=0.6\textwidth]{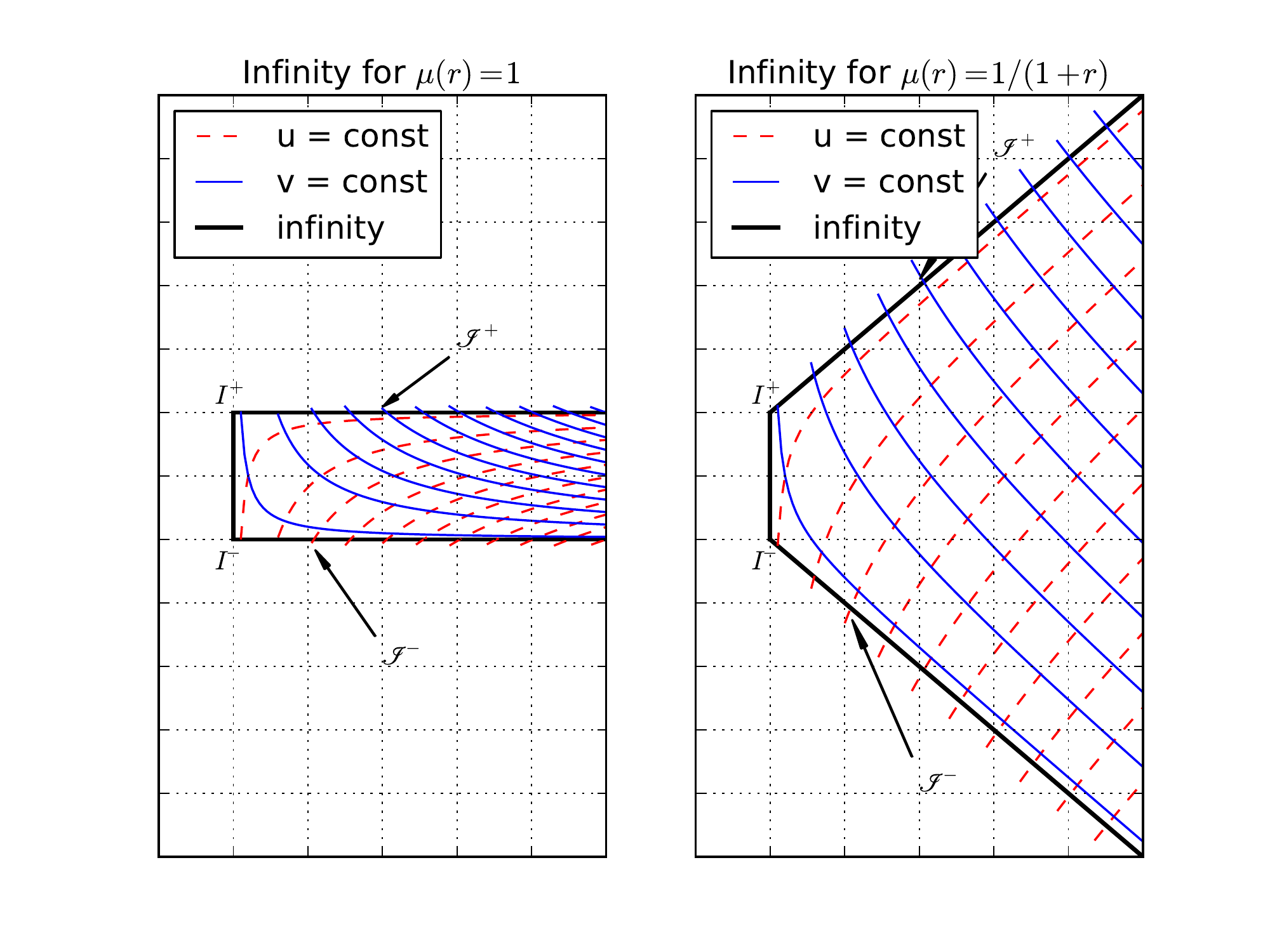}
  \caption{The neighborhood of the cylinder $I$ for the horizontal and the diagonal
    conformal representations. }
  \label{fig:cylinder}
\end{figure}

The physical metric of Minkowski space-time in Cartesian coordinates~$(y^\mu)$ reads
\begin{equation*}
  \tilde{g} = \eta_{\mu\nu}\, \dd y^\mu\, \dd y^\nu,
\end{equation*}
where $\eta_{\mu\nu} = \mbox{diag}(1,-1,-1,-1)$. In this representation the region that we 
are mainly interested in, namely the neighborhood of space-like infinity $i^0$, lies far away 
from the origin. In order to get a more detailed description of the points lying at infinity, 
we consider the coordinate inversion \cite{Penrose:1965p3324}: 
$y^\mu = -\frac{\mbox{x}^\mu}{\mbox{x}_\lambda\, \mbox{x}^\lambda}$ which brings our original 
metric into the form
\begin{equation*}
  \tilde{g} = \frac{1}{(\mbox{x}_\lambda\, \mbox{x}^\lambda)^2}\, \eta_{\mu\nu}\, \dd\mbox{x}^\mu\, 
  \dd\mbox{x}^\nu.
\end{equation*}
Notice that with the aforementioned 
inversion points close to infinity in our original coordinates $y^\mu$ are now lying in the 
vicinity of the origin of the new `inverted' coordinates $\mbox{x}^\mu$. But this comes at 
a price: the above metric is singular at the null-cone of the origin $\mbox{x}^\mu=0$. Introducing the conformal factor 
\begin{equation*}
  \Omega = - \mbox{x}_\lambda\, \mbox{x}^\lambda,
\end{equation*}
one can define the conformally equivalent metric 
\begin{equation*}  
  g' = \Omega^2 \tilde{g} = \eta_{\mu\nu}\, \dd\mbox{x}^\mu\, \dd\mbox{x}^\nu, 
\end{equation*}
which extends smoothly to the null-cone $\mbox{x}^\mu \mbox{x}_\mu = 0$. This implies that
space-like infinity is a regular point in Minkowski space-time. It is well-known that in
space-times with non-vanishing ADM-mass this is no longer the case and space-like infinity
can no longer be represented as a point. Instead, as discussed in detail
in~\cite{Friedrich:1998tc} it is more appropriate to represent the region near the origin as
a cylinder. This is achieved here by a further rescaling of the metric
\begin{equation*}
  g = \kappa^{-2} g'
\end{equation*}
with a function $\kappa(r) = r \mu(r)$ where $r = \sqrt{(\mbox{x}^1)^2 + (\mbox{x}^2)^2 +
  (\mbox{x}^3)^2}$ is the radial spatial distance from the origin and $\mu(r)$ is a
positive radial function with $\mu(0)=1$. The choice of the function $\mu$ represents the
remaining conformal gauge freedom. After the introduction of a new time-coordinate $t$ by
defining $x^0 = \kappa(r)t$ the metric $g$ expressed in polar coordinates takes the
following spherically symmetric form
\begin{equation}
 \label{cyl_metric}
  g = \kappa^{-2} \left(\kappa^2\mbox{d}t^2 + 2\, t\, \kappa\, \kappa'\mbox{d}t\, \mbox{d}r - 
  (1 - t^2\, \kappa'^2) \mbox{d}r^2 - r^2 (\mbox{d}\theta^2 + \sin^2 \theta\, \mbox{d}\phi^2) \right),
\end{equation}
where $(\theta, \phi)$ are the usual angular coordinates and ${}^\prime$ denotes differentiation 
with respect to the radial coordinate $r$. Notice that the final rescaled metric \eqref{cyl_metric} 
and the original Minkowski $\tilde{g}$ are related by the conformal factor
\begin{equation*}
 \Theta = \kappa^{-1}\, \Omega = \frac{r}{\mu(r)}\,(1 - t^2\, \mu(r)^2),
\end{equation*}
which is positive on $M = \{ r>0, |t| < 1/\mu(r)\}$, corresponding to the Minkowski
space-time. The conformal factor vanishes on the two regular 3-dimensional hyper-surfaces
\begin{equation*}
 \mathscr{I}^\pm = \{r>0,\, t = \pm\,\frac{1}{\mu(r)}\}\qquad
\text{and} \qquad I = \{r=0,\, |t| < 1\}.
\end{equation*}
The hyper-surfaces $\mathscr{I}^\pm$ are null, corresponding to null-infinity, while $I$
is a regular hyper-surface with the topology of a cylinder.  Interestingly, in the limit
$r \rightarrow 0$ future and past null-infinity do not meet at the same point as in the
conventional picture. Instead, they meet the cylinder $I$ in the 2-spheres $I^\pm = \{r=0,
t=\pm 1\}$. This observation leads naturally to the finite representation of space-like
infinity, where $i^0$ has been blown up to the cylinder $I = \{r=0, -1< t <1\}$. The
function $\mu$ dictates the shape of these structures. Here, the two simplest choices will
be considered: $\mu(r)=1$ and $\mu(r) = \frac{1}{1+r}$. In the former `horizontal'
representation, null-infinity is given by the set $\mathscr{I}^\pm = \{r>0,\, t =
\pm\,1\}$, while in the latter `diagonal' representation by the set $\mathscr{I}^\pm =
\{r>0,\, t = \pm(1+r)\}$, see Fig.~\ref{fig:cylinder}. The main advantage of working in the 
'horizontal' representation is that the entire space-time is contained between $t = \pm 1$, 
which allows us to study the whole space-time in a (formally) finite time evolution, i.e. reaching $t=1$. 
But, on the numerical level, working in this representation is more challenging, especially close 
to the region $I^+$ where the characteristic speed becomes infinite, see Figs.~\ref{fig:cylinder} 
and \ref{fig:charact_spin-2_wave}. Thus, the study of our system close to $I^+$ demands more 
computational resources and/or more elaborated time integration techniques---e.g. in \cite{Beyer:2012ti} 
we used an adaptive time step Runge-Kutta scheme to get as close as possible to $t=1$. For a further 
discussion on these and related issues see \cite{Beyer:2012ti, Doulis2012:phdthesis}.
It is helpful to introduce the double null coordinates 
\begin{equation}
u = \kappa(r) t - r, \qquad v = \kappa(r) t + r,
\end{equation}
which puts the metric into the form
\[
g = \frac1{\kappa^2}\dd u \dd v - \frac1{\mu^2}\dd\omega^2.
\]
Now, $\scri^\pm$ is characterized by the vanishing of exactly one of the null coordinates,
the other one being non-zero. Both coordinates vanish on the cylinder~$I$. In
Fig.~\ref{fig:cylinder} we also display the lines of constant $u$ and $v$. Since these are
coordinates adapted to the conformal structure of $M$ we can see clearly, that the
cylinder is `invisible' from the point of view of the conformal structure.  Notice, that
the differentiable structures defined by the $(u,v)$ and the $(t,r)$ coordinates are
completely different near the boundary~$r=0$.

\section{The spin-2 wave system}
\label{sec:spin-2_wave_system}

Our objectives in the present section are to derive a second order system for the spin-2 
equation described in~\cite{Beyer:2012ti} and to establish a correspondence between the
two formulations.

\subsection{The spin-2 wave equation}
\label{sec:relating_first_second_order_system}

Our starting point will be the spin-2 zero-rest-mass equation \cite{Penrose:1965p3324} on an
arbitrary space-time
\begin{equation}
 \label{spin-2_eq}
 \nabla^E{}_{A'} \phi_{EBCD} = 0,
\end{equation}
where $\phi_{ABCD}$ is a totally symmetric spinor-field on the space-time.
Now, we will derive a system of wave equations for $\phi_{ABCD}$ (see
eqn.~\eqref{spin-2_wave}) and show its equivalence
with~\eqref{spin-2_eq}. Establishing this correspondence is necessary because we need to know under
what conditions we obtain the same solutions. We will then also be able to compare the
numerical behavior of the two systems and study their similarities and differences. In
the following, we will prove that in a conformally flat space-time with vanishing Ricci
scalar a solution $\phi_{ABCD}$ of the spin-2
wave equation \eqref{spin-2_wave}, which satisfies initially the spin-2 zero-rest-mass equation
\eqref{spin-2_eq}, is also a solution of the latter. In the case of an initial-boundary
value problem (as we deal with in our numerical studies) the same statement is true
provided the boundary conditions for the wave system are determined from the first order system.

First, we define the spinor
\begin{equation}
 \label{defin_sigma_spinor}
  \Sigma_{A'BCD} \equiv \nabla^F{}_{A'} \phi_{FBCD}.
\end{equation}
Then the equation $\Sigma_{A'BCD} = 0$ is nothing else than the spin-2 equation
\eqref{spin-2_eq}. Taking another spinor derivative, we obtain an equation for the derivative
of $\Sigma_{A'BCD}$ 
\begin{equation} \label{eq:3}
 \nabla_{AA'} \Sigma^{A'}{}_{BCD} = \frac12 \Box \phi_{ABCD} + 3
\Psi_{A(B}{}^{EF} \phi_{CD)EF} + 6 \Lambda \phi_{ABCD} .
\end{equation}
Now, assume that $\phi_{ABCD}$ satisfies~\eqref{spin-2_eq}, then $\Sigma_{A'BCD}=0$ and 
\[
\frac12 \Box \phi_{ABCD} + 3\Psi_{A(B}{}^{EF} \phi_{CD)EF} + 6 \Lambda \phi_{ABCD} =0.  \]
Contracting over indices $A$ and $B$ yields \[ \Psi_{(C}{}^{AEF} \phi_{D)AEF} = 0, \] a
purely algebraic condition between the conformal curvature and the spinor field. This is a
Buchdahl condition~\cite{Penrose:1984wr} which implies the well-known fact that the
zero-rest mass equation (with spin 2 in this case) is inconsistent on space-times with
non-vanishing Weyl tensor. So we take $\Psi_{ABCD}=0$ from now on and we also assume that
$\Lambda=0$. Then the spinor field satisfies the second order wave equation 
\begin{equation}
\label{spin-2_wave}
  \Box\, \phi_{ABCD} = 0.
\end{equation}
Conversely, let $\phi_{ABCD}$ be a solution of~\eqref{spin-2_wave}. Then~\eqref{eq:3} shows
that the spinor field $\Sigma_{A'BCD} = 0$ satisfies the equation
\begin{equation}
 \nabla_{AA'} \Sigma^{A'}{}_{BCD} = 0.\label{eq:1}
\end{equation}
In order to study the properties of this system of PDE we look at its symbol. This is, for
every real co-vector $p_{AA'}$, defined as a map $P: \mathcal{S}^{A'}\otimes
\mathcal{S}_{(BCD)} \to \mathcal{S}_{A}\otimes \mathscr{S}_{(BCD)}$ between the
appropriate spin spaces defined by
\[
\alpha^{A'}\beta_{BCD} \mapsto p_{AA'} \alpha^{A'} \beta_{BCD}.
\]
Therefore, it is enough to discuss only the map 
\begin{equation}
p: \alpha^{A'} \mapsto p_{AA'} \alpha^{A'}.\label{eq:5}
\end{equation}
We define a sesqui-linear form on $\mathcal{S}^{A'}$ by
\[
\left<\beta,\alpha\right>:= \bar\beta^Ap_{AA'}\alpha^{A'}, \qquad \text{for all }\alpha^{A'},
\beta^{A'} \in \mathcal{S}^{A'}.
\]
It is easy to verify that $\left<\cdot,\cdot\right>$ is Hermitian due to the reality of
$p_{AA'}$, i.e.,  
\[
\overline{\left<\beta,\alpha\right>} = \left<\alpha,\beta\right>.
\]
Furthermore,
\[
\left<\alpha,\alpha\right> = \bar\alpha^Ap_{AA'}\alpha^{A'} 
\]
is positive definite, if and only if $p_{AA'}$ is a time-like and future-pointing
covector. Since the
map $P$ simply consists of four copies of the map $p$ this result shows that the full
system~\eqref{eq:1} is symmetric hyperbolic and that, consequently, its Cauchy problem is
well-posed. 

Thus, if $\Sigma_{A'BCD}=0$ on an initial hyper-surface and, for initial boundary value
problems, if boundary data are given such that the ingoing  $\Sigma_{A'BCD}$ waves vanish,
then the field $\Sigma_{A'BCD}$ vanishes everywhere, i.e., the field $\phi_{ABCD}$
satisfies \eqref{spin-2_eq}. This conclusion holds on the region where the systems are
hyperbolic. In our application below we will consider situations where hyperbolicity is
lost on a region with codimension 2, namely the spheres $I^\pm$. We find empirically that the
conclusion still holds near these regions.

We will use this result as follows. In the numerical implementation of \eqref{spin-2_wave}
the initial data and the boundary conditions are determined from \eqref{spin-2_eq}. Then,
the computed solution will be compared to a previously computed \cite{Beyer:2012ti} solution 
of \eqref{spin-2_eq}.

\subsection{Coordinate representation of the second order wave equation}
\label{sec:coord-repr-second}

In order to obtain a coordinate representation of \eqref{spin-2_wave}, we 
introduce a null-tetrad that is adapted to the spherical symmetry of the background metric~\eqref{cyl_metric}. We will choose the same 
null tetrad $(l^\nu, n^\nu, m^\nu, \bar{m}^\nu)$ as in~\cite{Beyer:2012ti}, namely 
\begin{equation*}
 \begin{aligned} 
  l^\nu &= \frac{1}{\sqrt{2}}(1 - t \kappa',\kappa,0,0),  
  \qquad n^\nu = \frac{1}{\sqrt{2}}(1 + t \kappa',-\kappa,0,0), \\
  m^\nu &= \frac{\mu}{\sqrt{2}}\left(0,0,1,-\frac{\mathrm{i}}{\sin\theta}\right), 
  \qquad \bar{m}^\nu = \frac{\mu}{\sqrt{2}}\left(0,0,1,\frac{\mathrm{i}}{\sin\theta}\right).
 \end{aligned}
\end{equation*}
The non-vanishing spin-coefficients for this tetrad are
\begin{equation}
  \label{eq:spin-coeffs}
\rho=-\rho' = \frac1{\sqrt2} r \mu', \qquad \epsilon=\gamma= -\frac1{2\sqrt2} \kappa'.
\end{equation}
Now we express the wave operator $\square$ in terms of the weighted differential 
operators of the GHP (Geroch-Held-Penrose) formalism \cite{Penrose:1984wr} using
the commutation relations
\begin{equation*}
 \begin{aligned}
  & (\mbox{\th}'\mbox{\th} - \mbox{\th}\mbox{\th}') \phi_{ABCD} = 0, 
  \qquad (\eth \mbox{\th} - \mbox{\th} \eth) \phi_{ABCD} = - \rho \, \eth \phi_{ABCD}, \\
  & (\eth \mbox{\th}' - \mbox{\th}' \eth) \phi_{ABCD} =  -\rho' \, \eth' \phi_{ABCD}, 
  \qquad (\eth' \eth - \eth \eth') \phi_{ABCD} = 0,
 \end{aligned}
\end{equation*}
which contain only the non-vanishing spin-coefficients. Expanding the totally symmetric 
spin-2 zero-rest-mass field in the familiar way 
\begin{equation*}
 \begin{aligned}
  \phi_{ABCD} \equiv \iota_{A} \iota_B \iota_C \iota_{D} \phi_0 &- 4\,\iota_{(A} \iota_B \iota_C o_{D)} \phi_1 +\\ 
  &+6\,\iota_{(A} \iota_B o_C o_{D)} \phi_2 - 4\,\iota_{(A} o_B o_C o_{D)} \phi_3 + o_{A} o_B o_C o_{D} \phi_4,
 \end{aligned}
\end{equation*}
we obtain the following system of five equations
\begin{equation}
 \label{spin-2_wave_GHP}
  \mbox{\th}\mbox{\th}' \phi_k - \rho' \, \mbox{\th} \phi_k - 
  \rho \, \mbox{\th}' \phi_k + k (5 - k) \, \rho^2 \, \phi_k = 
  \eth\eth' \phi_k + (4 - k) \, \rho \, \eth\phi_{k+1} - 
  k \, \rho \, \eth'\phi_{k-1} 
\end{equation}
with $k = 0, 1, 2, 3, 4$.

Expressing the GHP operators $\mbox{\th}$ and $\mbox{\th}'$ in terms of the directional
derivatives along the tetrad vectors $l$ and $n$ (see~\cite{Penrose:1984wr}) yields
\begin{equation*}
  \begin{aligned}
   &\mbox{\th}\, \eta = \frac{1}{\sqrt{2}}\, ((1 - t \kappa')\, \partial_t + \kappa\, \partial_r - 
   2\,\sqrt{2}\, w\, \epsilon)\, \eta, \\ 
   &\mbox{\th}' \eta = \frac{1}{\sqrt{2}}\, ((1 + t \kappa')\, \partial_t - \kappa\, \partial_r - 
   2\,\sqrt{2}\, w\, \gamma)\, \eta,
  \end{aligned}
\end{equation*}
where $\eta$ is a $\{p, q\}$-scalar quantity with boost-weight $w=\frac{p+q}{2}$ and
spin-weight $s=\frac{p-q}2$, see~\cite{Penrose:1984wr}. Due to the spherical symmetry of the
metric and the adapted null-tetrad it is natural to refer the $\eth$ and $\eth'$ operators
to the unit-sphere, so that we replace
\begin{equation*}
 \eth \mapsto \frac{\mu}{\sqrt{2}}\, \eths, \quad 
 \eth' \mapsto \frac{\mu}{\sqrt{2}}\, \eths'
\end{equation*}
in~\eqref{spin-2_wave_GHP} as in \cite{Beyer:2012ti}. (Where with $\eths, \eths'$ we denote 
the edth operators on the unit-sphere.) Then the system \eqref{spin-2_wave_GHP} takes the form
\begin{equation}
 \label{spin-2_wave_coord}
  \begin{aligned}
    (1 - t^2 \kappa'^2) &\partial_{tt}\phi_k - \kappa^2 \partial_{rr}\phi_k + 2 \, t \kappa
    \kappa' \partial_{tr}\phi_k \\
    &+ 2 \, [(2 - k) \kappa' - t (\kappa'^2 +
    r\, \mu' \kappa' - \frac{1}{2} \kappa \, \kappa'')] \partial_{t}\phi_k + 2 \, r \kappa
    \mu' \partial_{r}\phi_k \\
    &+ [(2 - k) (\kappa \kappa'' + (1 - k) \kappa'^2) +k (5 - k) r^2 \mu'^2] \phi_k \\
    & \hspace*{4em}
    = \mu^2 \, \eths\eths' \phi_k + (4 - k) \, r \, \mu \, \mu' \, \eths\phi_{k+1} - 
   k \, r \, \mu \, \mu' \, \eths'\phi_{k-1}.
  \end{aligned}
\end{equation}
Finally, we use the spherical symmetry of the metric \eqref{cyl_metric} to expand the
components $\phi_k$ of the spin-2 field as a sum of spin-weighted spherical harmonics
${}_s Y_{lm}$ in the following way 
\begin{equation*}
 \label{expan}
  \phi_k(t, r, \theta, \phi) = \sum_{lm} \phi_k^{lm}(t,r)\, {}_{2-k} Y_{lm}(\theta, \phi),
\end{equation*}
where $s=2-k$ is the spin-weight of $\phi_k$ and the integers $s, l, m$ satisfy the inequalities 
$|s|\leq l$ and $|m|\leq l$. Since the operators $\eths, \eths'$ act on the
spin-weighted spherical harmonics $_s Y_{lm}$ as
\begin{equation*}
  \begin{aligned}
    \eths(_s Y_{lm}) &= - \sqrt{l(l+1)-s(s+1)}\; {}_{s+1} Y_{lm}, \\
    \eths'(_s Y_{lm}) &= \sqrt{l(l+1)-s(s-1)}\;  {}_{s-1} Y_{lm}, \\
    \eths\eths'(_s Y_{lm}) &= - (l(l+1)-s(s-1))\; {}_s Y_{lm},
  \end{aligned}
\end{equation*} 
the system~\eqref{spin-2_wave_coord} decouples into separate systems for each
mode of the fixed pair $(l, m)$, i.e.
\begin{equation}
 \label{spin-2_wave_final}
  \begin{aligned}
    (1 - t^2 \kappa'^2) \partial_{tt}\phi_k &- \kappa^2 \partial_{rr}\phi_k + 2 \, t \kappa
    \kappa' \partial_{tr}\phi_k \\
    &+ 2 \, [(2 - k) \kappa' - t (\kappa'^2 +
    r\, \mu' \kappa' - \frac{1}{2} \kappa \, \kappa'')] \partial_{t}\phi_k + 2 \, r \kappa
    \mu' \partial_{r}\phi_k \\
    &+ [(2 - k) (\kappa \kappa'' + (1 - k) \kappa'^2) +k (5 - k) r^2 \mu'^2] \phi_k \\
    & \hspace*{3em}
    = - \mu^2 \, c^2_k \phi_k - r\, \mu \, \mu' (
   (4 - k) c_k \,  \phi_{k+1} +  k \, c_{k-1} \, \phi_{k-1}) ,
  \end{aligned} 
\end{equation}
where $c_k = \sqrt{l(l + 1) - (2-k)(1-k)}$ whenever the square root is real and otherwise
$c_k=0$. For the sake of simplicity the notation $\phi_k^{lm}=\phi_k$ was introduced. It
is worthwhile to mention that this form of the system holds for $l\ge2$, since for $l=1$
the components $\phi_0$ and $\phi_4$ vanish and for $l=0$ only $\phi_2$ survives. Since
the above system comes from equations involving the wave operator it is hyperbolic.

\subsection{Characteristic curves}
\label{sec:characteristics}

Because of its hyperbolic nature the system \eqref{spin-2_wave_final} has two families of
real characteristic curves. As in the case of the spin-2 equation, their behavior will be
very useful in the subsequent numerical studies. Following \cite{John:1982}, the slope of
the characteristic curves for the system \eqref{spin-2_wave_final} is given by
\begin{equation*}
 \frac{dt}{dr} = \pm\frac{1 - t \kappa'(r)}{\kappa(r)}.
\end{equation*}
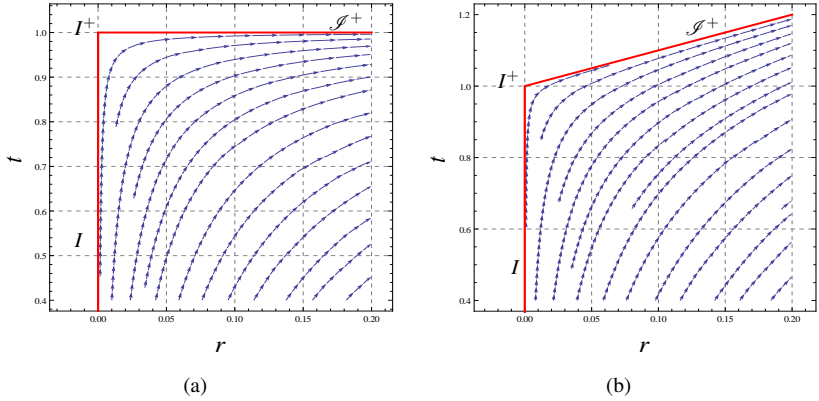
\begin{figure}[htb]
  \centering
  \subfigure[]{\includegraphics[height=5.cm,width=5.5cm]{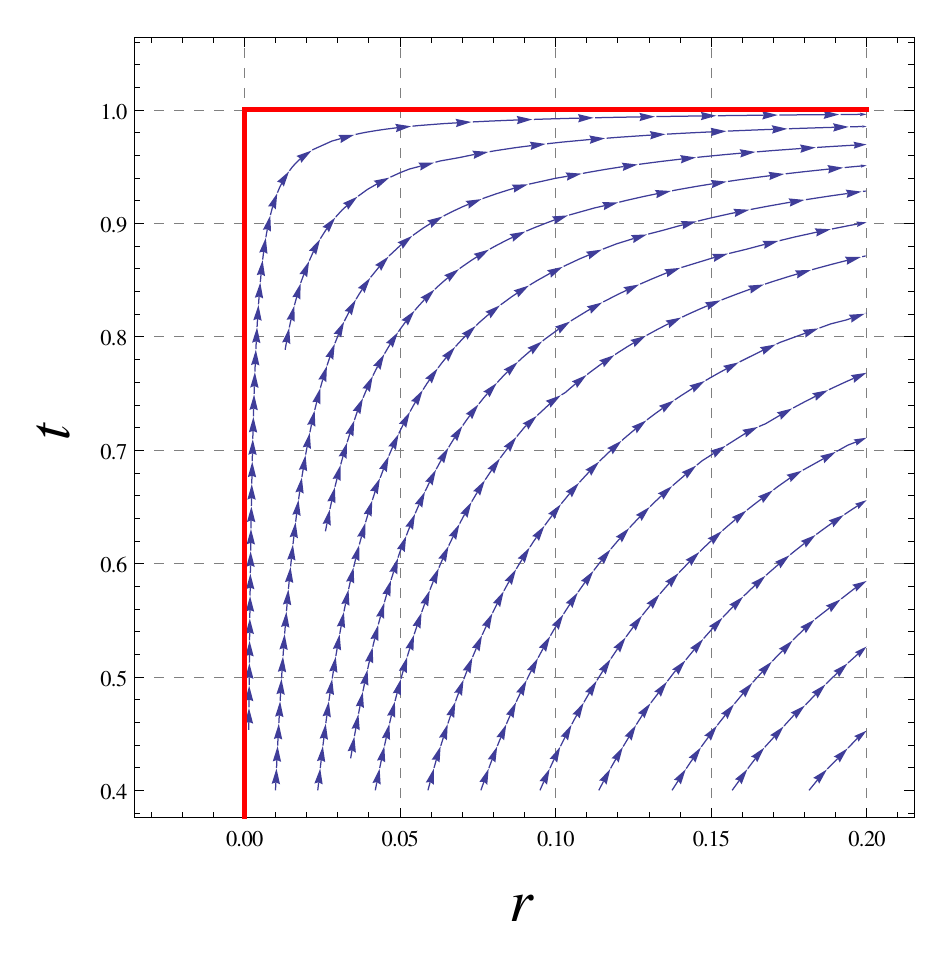}
  \put(-30,127){$\mathscr{I}^+$}
  \put(-125,125){$I^+$}
  \put(-125,45){$I$}}
  \subfigure[]{\includegraphics[height=5.cm,width=5.5cm]{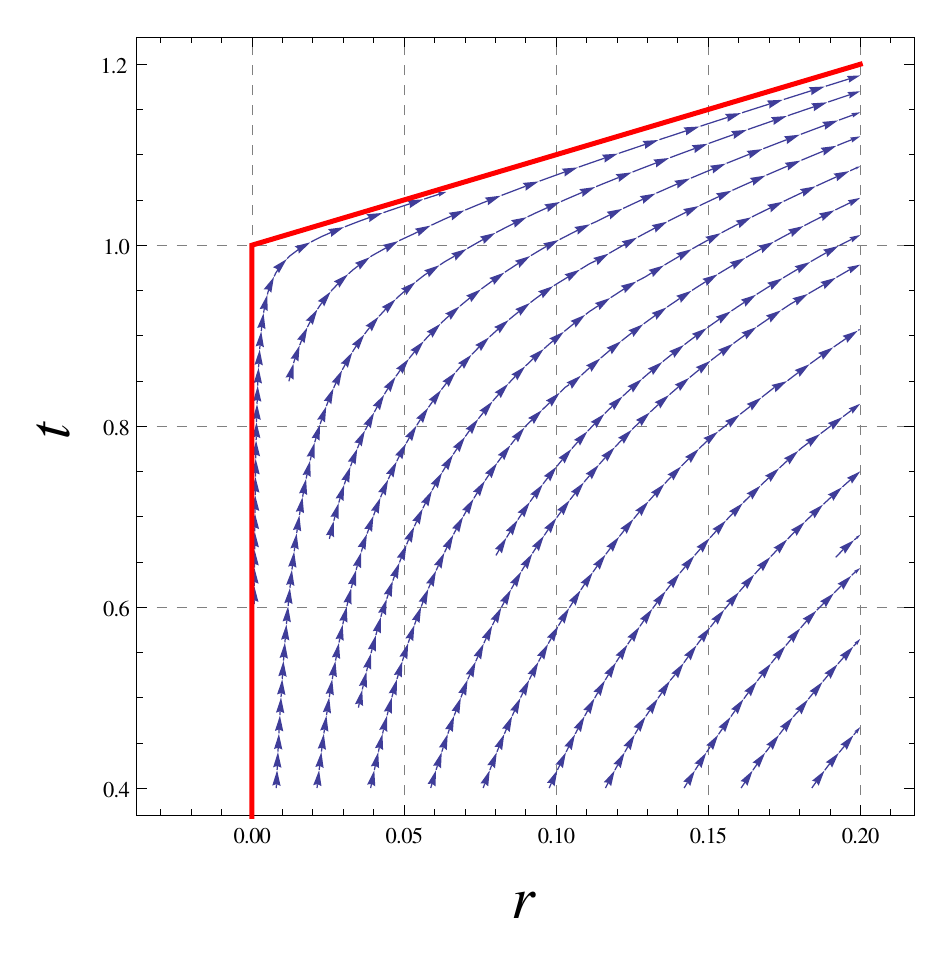}
  \put(-55,125){$\mathscr{I}^+$}
  \put(-125,105){$I^+$}
  \put(-120,35){$I$}}
  \caption{\label{fig:charact_spin-2_wave}
    Characteristic curves of all the fields $\phi_k$
  in a neighborhood of $I^+$. The red line denotes the cylinder and null-infinity. 
  The situation close to $I^-$ is obtained by a reflection in the 
  $r$-axis. (a) Characteristic curves of slope $\frac{dt}{dr} = 
  \frac{1 - t}{r}$ in the `horizontal' case $\mu=1$. (b) Characteristic curves 
  of slope $\frac{dt}{dr} = \frac{1 - t \kappa'}{\kappa}$ in the `diagonal' case $\mu = 
  \frac{1}{1 + r}$.}
\end{figure}%
The characteristic curves of the spin-2 wave equation are identical 
to those of the spin-2 equation. Fig.~\ref{fig:charact_spin-2_wave} shows the 
`outgoing' characteristic curves of \eqref{spin-2_wave_final} with slope 
$dt/dr = (1 - t \kappa')/\kappa$ near a neighborhood of $I^+$; near $I^-$ these 
curves tend to become parallel to the cylinder $I$. On the other hand, the `ingoing' 
characteristic curves with slope $dt/dr =- (1 + t \kappa')/\kappa$ are becoming 
parallel to $I$ near $I^+$; their behavior in a neighborhood of $I^-$ can be 
visualized by a reflection of Fig.~\ref{fig:charact_spin-2_wave} along the $r$-axis.
 
There is a slight difference with the first order case, though. Now, the behavior of the
characteristic curves is universal, in the sense that all the components of the spin-2
field propagate along the same characteristics, while in the first order system different
components have in general different characteristics (depending on the gauge). 

Furthermore, as in the first order system the evolution equation reduce to an interior
system on the cylinder in the sense that there remain no $r$-derivative terms in the
equations when evaluated at $r=0$. So, as before, the cylinder is a total characteristic
for the system. And as before, the equations loose their hyperbolicity at $I^\pm$, since
the coefficient in front of the second time derivative vanishes there.

\section{Numerical analysis and results}
\label{sec:numerical_analysis}

In this section, we will numerically solve the initial boundary value problem 
for the spin-2 wave equation \eqref{spin-2_wave_final} using the findings of 
sec.~\ref{sec:relating_first_second_order_system}. Specifically, we will prescribe 
initial data that satisfy the constraints of the first order system and subsequently 
evolve them with \eqref{spin-2_wave_final}. In addition, the boundary conditions 
will be prescribed using the first order system.

\subsection{Numerical implementation}
\label{sec:numerical_implementation}

We discretize the $1 + 1$ system \eqref{spin-2_wave_final} using the method of
lines. Accordingly, the system \eqref{spin-2_wave_final} is reduced to a system of
ordinary differential equations by discretizing the spatial coordinate $r$. In this work,
we use finite difference techniques to achieve that.

As computational domain we adopt the interval $D = [0,1]$. A finite representation of $D$
is obtained through the introduction of an equidistant grid $r_i = i h$ with
$i=0,\ldots,N$ on $D$, where $r_N = 1$, and $h$ is the grid spacing. The spatial
derivative operators appearing in \eqref{spin-2_wave_final} are approximated by
appropriate \emph{summation by parts} (SBP) finite difference operators
\cite{Strand:1994ef,Carpenter:1999cl,Mattsson:2004br,Diener:2007bx}. The SBP operators
arise from discrete versions of the continuous energy estimates that were originally
introduced in~\cite{Kreiss:1974ti,Kreiss:1977}. While these energy estimates guarantee the
well-posedness in the continuous case, on the discrete level their discrete counterparts
guarantee the numerical stability of our discrete schemes. This is the main reason we
chose to use these specific finite difference operators. 

Another very appealing property of the SBP operators is that although their accuracy near
the boundaries is, depending on the details of their construction, one or two orders
smaller than the one in the interior of the grid, the overall accuracy is of the same
order with the accuracy in the interior. 

We approximate the second order derivatives with the minimal width, full norm SBP operator
given in \cite{Mattsson:2004br}, which is fourth order accurate in the interior and second
order at the boundary. We call an SBP operator with these properties a
$(4,2)$-operator. The first order derivatives are approximated by the corresponding first
order SBP operator with the same norm matrix originally constructed in
\cite{Carpenter:1999cl}, which is a $(4,3)$-operator. The results presented below were
obtained with the above choice of SBP operators.

We also experimented with SBP operators based on a diagonal norm; specifically, the second
derivatives were approximated by a minimal width, diagonal norm $(4,2)$-operator and the
first derivatives by a $(4,2)$-operator with the same norm, both given in
\cite{Mattsson:2004br}. This combination of operators is expected (see
\cite{Mattsson:2004br}) to have a third order overall accuracy. Our numerical findings,
see \cite{Doulis2012:phdthesis}, confirmed this expectation.

In addition, we combined the above full norm second order SBP operator with two first
order SBP operators with different norm constructed in \cite{Strand:1994ef,
  Diener:2007bx}, which were very successfully used already in the numerical
implementation of the first-order formulation of the spin-2 equation, see
\cite{Beyer:2012ti, Doulis2012:phdthesis}. Namely, the first order derivatives were
approximated with restricted full norm $(4,3)$-operators. Of course, with this choice we
violate the assumption made in~\cite{Mattsson:2004br}, of combining first and second order SBP
operators of the same norm. Theoretically though, under certain conditions on the norm
matrices of the two operators, energy estimates can still be obtained.  So, according to
our numerical findings \cite{Doulis2012:phdthesis}, it seems that requiring the same norm
for first and second order operators is sufficient but not necessary for stability.

A point that also needs special attention is the imposition of the boundary conditions. A
wrong imposition of the boundary condition would destroy the designed accuracy of the SBP
operators and lead to instabilities \cite{Carpenter1994TimeStable}. We use a very simple,
but highly efficient, \textit{penalty} method---the so-called \emph{simultaneous
  approximation term} (SAT) method introduced in
\cite{Carpenter1994TimeStable}---that preserves the designed accuracy of the SBP operators and
guarantees the numerical stability of our schemes. The combination of the SBP operators
with the SAT method has been successfully applied in several different circumstances, see
e.g.,~\cite{Carpenter:1999cl,Mattsson:2004br,Diener:2007bx,Carpenter1994TimeStable,Lehner:2005hc,Schnetter:2006ku,Gong:2011gr}.

Finally, one has to decide how to solve the semi-discrete system of ODEs obtained from
\eqref{spin-2_wave_final} after the spatial discretization. Here, we implement the system
as a first-order system of ODEs by introducing the first order time derivatives of the
grid functions as additional variables. This system is then solved using the standard 
4th order Runge-Kutta scheme. The code was written from scratch in Python.

\subsection{Treatment of the boundaries}
\label{sec:boundaries}

As it was already mentioned in sec.~\ref{sec:characteristics}, the cylinder is a total
characteristic of the system \eqref{spin-2_wave_final}. Thus, we do not have to prescribe
boundary conditions for the points of our computational domain that lie on the cylinder,
i.e. at $r=0$. In contrast, at the boundary $r=1$ there is one ingoing characteristic for
each component of the spin-2 field. Thus, we have to provide appropriate boundary
conditions at $r=1$ for each component $\phi_k$. But, in order to compute the same
solution as with the first order system we need to use all the available information
from the first-order system to determine boundary conditions for the second order
system. Essentially, we require the first order system to hold not only on the initial
surface but also on the boundaries. Then, there remains just one free function, namely
$\phi_0(t,1)$, which can be specified arbitrarily on the boundary 
$r=1$. (For a further analysis on this issue see \ref{sec:general_data}.) To implement the
boundary conditions we can use different formulations of boundary 
conditions which are known to lead to well-posed problems for the wave equation: Neumann or, more 
generally, Robin conditions of the form 
\begin{equation}
 \label{Robin_boundary}
  \phi_k(t, 1) + \partial_r \phi_k(t, 1) = g_k(t),
\end{equation}
where $k = 0,\ldots,4$, as well as Sommerfeld type boundary conditions. Since we need to
use informations from the 1st order formulation of the spin-2 equation (as discussed in
sec.~\ref{sec:relating_first_second_order_system}), which provides relations for the
values of the first derivatives of the components but not for the values of the functions
themselves, we cannot impose Dirichlet conditions.

To motivate the above choice of boundary conditions and to exemplify the use of the SAT method for 
systems with second derivatives, we will consider the simplest form of a 1-D wave equation 
\begin{equation}\label{eq:4}
 u_{tt} = u_{xx}, \qquad 0 \leq x \leq 1, \quad t \geq 0
\end{equation}
with homogeneous Robin boundary conditions
\begin{equation*}
  \alpha\, u(t,0) + u_x(t, 0) = 0, \qquad \beta\, u(t, 1) + u_x(t, 1) = 0,
\end{equation*}
where $\alpha, \beta$ are arbitrary constants. The energy method (i.e. multiplying \eqref{eq:4} 
by $u_t$ and integrating by parts) for the above boundary value problem gives 
\begin{equation*}
  \frac{d}{dt}\left(\int^1_0 (u_t^2 + u_x^2)\,dx \right) = \frac{d}{dt} \left(-\beta\, u^2(t, 1) + \alpha\, u^2(t, 0)\right),
\end{equation*}
which leads to an energy estimate for $\beta \geq 0$ and $\alpha \leq 0$. Thus, we have 
to choose $\beta \geq 0$ and $\alpha \leq 0$.

The semi-discrete approximation of~eq.~\eqref{eq:4} is $\upsilon_{tt} = D_2\upsilon$,
where $\upsilon = (\upsilon_i)_{i=0:N}$ is the grid function approximating the solution
$u$ and the $(N+1) \times (N+1)$ matrix $D_2$ is the SBP operator approximating the second
derivative operator $\partial^2/\partial x^2$. We define $D_2=H^{-1}(-A + B\, S)$, see \cite{Mattsson:2004br} 
for the details, where $A + A^T \geq 0$ is positive semi-definite, $H$ is the norm matrix, $S$ 
is an approximation of the first order derivative operator at the boundary and 
$B = \mathrm{diag}(-1,0,\dots,0,1)$.\footnote{All the above are $(N+1) \times (N+1)$ matrices.} 
Following \cite{Mattsson:2004br,Mattsson:2012}, the implementation of the boundary conditions with 
the SAT method leads to the introduction of a couple of boundary terms in the following fashion 
\begin{equation*}
  \upsilon_{tt}= H^{-1}(-A + B\, S) \upsilon + \tau_0\,H^{-1} E_0 (\alpha\,I + S) \upsilon 
  + \tau_N \,H^{-1} E_N (\beta\,I + S) \upsilon,
\end{equation*}
where $E_0 = \mathrm{diag}(1,0,\ldots,0)$, $E_N = \mathrm{diag}(0,\ldots,0,1)$, $I =
\mathrm{diag}(1,\ldots,1)$ are matrices of size $(N+1) \times (N+1)$, and $\tau_0 , \tau_N$ are the so-called penalty parameters. By choosing $\tau_0=1, \tau_N =-1$ the last expression simplifies considerably 
\begin{equation}
 \label{wave_SAT}
  \upsilon_{tt}= - H^{-1} (A + C)\upsilon,
\end{equation}
where $C (= C^T) \equiv - \alpha E_0 + \beta E_N$ is positive semi-definite if $\beta \geq 0$ and 
$\alpha \leq 0$. 

In~\cite{Mattsson:2004br} it was pointed out that for symmetric $A = A^T$ the energy method (i.e., multiplying \eqref{wave_SAT} from the left by $\upsilon^T_t H$ and adding to it the transpose of the resulting expression, see \cite{Mattsson:2009,Mattsson:2012}) yields
\begin{equation*}
 \frac{d}{dt}\left(\upsilon^T_t H \upsilon_t + \upsilon^T A \upsilon \right) = 
 \frac{d}{dt} (\alpha \upsilon_0^2 - \beta \upsilon_N^2), 
\end{equation*}
which leads to an energy estimate when the conditions $\beta \geq 0,\, \alpha \leq 0$ are satisfied.

But, as it was already mentioned in sec.~\ref{sec:numerical_implementation}, we also use a second order SBP operator with a non-symmetric matrix $A$, so that there exists no (straightforward) energy estimate. However, we have numerically checked that the matrix $-H^{-1} (A + C)$ has strictly negative real eigenvalues and is diagonalisable. This implies that \eqref{wave_SAT} has bounded solutions. Our simulations based on this operator are stable.

This procedure for imposing boundary conditions holds true \cite{Mattsson:2009} also for Neumann and
Sommerfeld conditions, but not for Dirichlet conditions. In that case, as was shown in
\cite{Mattsson:2009}, the method of imposing boundary conditions with the SAT method must
be modified in order to guarantee the stability of the numerical scheme. As discussed 
above, we do not have the need for imposing Dirichlet conditions. Therefore, by choosing \eqref{Robin_boundary}, Neumann or Sommerfeld boundary
conditions, we do not have to go into these additional complications.

\subsection{The exact solution}
\label{sec:exact_solution}

As described in sec.~\ref{sec:spin-2_wave_system}, the system \eqref{spin-2_wave_final} is just a second 
order reformulation of the spin-2 equation. Therefore, it must be satisfied by the exact solution presented 
in \cite{Beyer:2012ti,Doulis2012:phdthesis}:
\begin{equation}
 \label{exact}
  \begin{aligned}
    \phi_0(t,r) &= \frac{r^2 (1+r - t)^4}{(1+r)^7}, \qquad
    \phi_1(t,r) = \frac{2 r^2 (1+ r - t)^3 (1+ r + t)}{(1 + r)^7}, \\
    \phi_2(t,r) &= \frac{\sqrt{6} r^2 (1+ r - t)^2 (1+ r + t)^2}{(1+ r)^7}, \\
    \phi_3(t,r) &= \frac{2 r^2 (1+ r - t) (1+ r + t)^3}{(1+ r)^7}, \qquad
    \phi_4(t,r) = \frac{r^2 (1+ r + t)^4}{(1+ r)^7}.    
  \end{aligned}
\end{equation}
Recall that this solution refers to the $l=2$ mode in the `diagonal' representation of null-infinity. 
According to the discussion in sec.~\ref{sec:relating_first_second_order_system}, the initial data 
must satisfy the constraints of the first order system and be subsequently evolved with the second 
order system \eqref{spin-2_wave_final}. Therefore, evaluating \eqref{exact} at $t=0$, the initial 
data for each one of the fields are
\begin{equation}
 \label{initial_data_1}
   \phi_0(r) = \phi_4(r) = \frac{r^2}{(1+r)^3}, \quad
   \phi_1(r) = \phi_3(r) = \frac{2\, r^2}{(1 + r)^3}, \quad
   \phi_2(r) = \frac{\sqrt{6}\, r^2}{(1+ r)^3}\,.
\end{equation}
In a similar way, i.e. differentiating \eqref{exact} with respect to $t$ and evaluating at $t=0$, 
the first temporal derivative of the components of the spin-2 field read
\begin{equation}
 \label{initial_data_2}
 \dot\phi_0(r) = \dot\phi_1(r) = - \dot\phi_3(r) = 
   -\dot\phi_4(r) = - \frac{4\,r^2}{(1+r)^4}, \quad
   \dot\phi_2(r) = 0.
\end{equation}
In summary, \eqref{initial_data_1} and \eqref{initial_data_2} will be our initial conditions in 
this section. In addition, at $r=1$, we have to prescribe Robin boundary conditions of the form 
\eqref{Robin_boundary}. By differentiating \eqref{exact} with respect to the spatial coordinate 
$r$ and evaluating at $r = 1$ we obtain 
\begin{equation*}
  \begin{aligned}
   g_0(t) &= - \frac{1}{256}\, (t - 2)^3 (6 + t), \quad 
   g_1(t) = \frac{1}{128}\, (48 - 32 t + t^4),\quad \\
   g_2(t) &= - \frac{1}{128}\, \sqrt{\frac{3}{2}}\, (t^2 - 4) (12 + t^2),\quad \\
   g_3(t) &= \frac{1}{128}\, (48 + 32 t + t^4), \quad 
   g_4(t) = - \frac{1}{256}\, (t - 6) (2 + t)^3,   
  \end{aligned}
\end{equation*}
where the subscript denotes the component of the spin-2 field to which each boundary condition 
corresponds. The above boundary conditions are imposed with the SAT method.

\begin{table}[htb]
 \centering
  \begin{tabular}{|c|cc|ccc|}
    \hline
          &    \qquad \qquad $\phi_0$        & &   \qquad\qquad\qquad$\phi_4$ &&\\
    \hline
    Grid  &      1st order     &   2nd order  &    1st order     & 2nd order   &\\
    \hline
    50    &     -25.2218       &   -27.6006   &    -11.1643      &  -12.3418   &\\
    \hline 
    100   &     -29.3956       &   -31.5941   &    -13.9924      &  -15.1743   &\\ 
    \hline  
    200   &     -33.7109       &   -35.6075   &    -16.9978      &  -18.1782   &\\ 
    \hline
    400   &     -38.1000       &   -39.6068   &    -20.1075      &  -21.2850   &\\
    \hline
  \end{tabular}
  \caption{The logarithm of the normalized $l^2$ norm of the absolute error $E$,
    $\log_2(||E||_2)$, between the exact solution and the solutions computed from the
    1st-order system~\eqref{spin-2_eq} and the 2nd-order system~\eqref{spin-2_wave} at
    time $t=1$. Note, the 4th order convergence in $\phi_0$ and the 3rd-order convergence
    in $\phi_4$. 
  }
  \label{tab:compar_exact}
\end{table}
Evolving the initial data \eqref{initial_data_1} and \eqref{initial_data_2} with the second 
order system \eqref{spin-2_wave_final} we can reach $t=1$ without loss of the expected 4th
order convergence. In addition, the constraints are preserved during the
evolution. The above results strongly indicate that our code reproduces successfully the exact solution
\eqref{exact}. 

But the purpose of this work is to compare the present numerical approach with the one developed 
in \cite{Beyer:2012ti}, thus a first step towards this goal would be to check which approach 
reproduces better the exact solution \eqref{exact}.  Tab.~\ref{tab:compar_exact} serves this 
purpose. Specifically, it depicts the logarithm of the normalized $l^2$ norm of the absolute error 
$E$, i.e. $\log_2(||E||_2)$, of the components $\phi_0, \phi_4$ for the two numerical approaches 
at time $t=1$. Notice that a slightly better accuracy is achieved in the second order approach; a 
result that confirms the first of the claims made in \cite{Kreiss:2002du}.

We have also done the comparisons in the `horizontal' case with similar results. Here, we
cannot reach $I^+$ due to the degeneracy of the equation at $t=1$ which affects the
numerical algorithm because the propagation speeds grow unboundedly. However, we can come
arbitrarily closely to $\scri^+$ using an adaptive time-step. We will not go into any
further details here because the phenomena are the same as discussed in~\cite{Beyer:2012ti}.

\subsection{General initial data}
\label{sec:general_data}

According to the results of the preceding section, there is strong evidence that our
numerical code converges with the expected order and reproduces successfully the exact
solution \eqref{exact} of the system \eqref{spin-2_wave_final}. With the confidence that
these results provide us, we can proceed further in the numerical study of the spin-2 wave
system and seek for numerical solutions, which do not correspond to exact solutions of the
system~\eqref{spin-2_wave_final}. In addition, we will confirm numerically the analytic
result of sec.~\ref{sec:relating_first_second_order_system}, namely that the solutions of
the two different formulations for the spin-2 equation coincide if the initial and boundary
data are chosen appropriately.

Again, the initial data satisfy the constraints of the 1st order system and are evolved
with the second order system \eqref{spin-2_wave_final}. We use the constraints in the form
derived in sec.~3.3 of \cite{Beyer:2012ti} to determine the initial values of all the
components in terms of one free function. We define the auxiliary variables $\psi_i(r) :=
\phi_i(r)/\mu(r)^3$ for which the constraints imply the following relationships
\begin{equation}
  \label{constraints}
  \begin{aligned}
    \psi_0(r) = \psi_4(r) &= \frac{-\alpha_0^2 \psi_2(r) + 2\, r\, \psi'_2(r) +
      2\, r^2\, \psi''_2(r)}{\alpha_0\, \alpha _2}, \\
    \psi_1(r) = \psi_3(r) &= \frac{r\, \psi'_2(r)}{\alpha_0},
  \end{aligned}
\end{equation}
with $\alpha_0=\sqrt{l(l+1)}$ and $\alpha_2=\sqrt{l(l+1)-2}$.

This choice leaves the component $\phi_2$ completely at our disposal and allows us to
compute the remaining components explicitly. Note, that this is not the
most general form of the initial data. We have restricted ourselves to the case where
$\phi_3(r)=\phi_1(r)$. Otherwise there would exist another free function.

We take the initial values of $\phi_2$ in the form of a bump function
\begin{equation}
  \label{algebr_ic_phi_2}
  \phi_2(r) = 
    \begin{cases} 
      \left(4\,\frac{r}{b} (1 - \frac{r}{b})\right)^{16}, & 0 \leq r \leq b\\
      0, & b \leq r \leq 1
    \end{cases}.
\end{equation}
centered at $r=b/2$, then all the other components are bump functions as well. In this
section, we will choose $b = 1$. 

We also have to specify initially the first time derivatives of the components. The
evolution equations of the first order system, see Appendix~\ref{sec:first-order-equat}, 
will be used for this purpose. Evaluated at $t=0$ they read 
\begin{equation}
  \label{algebr_ic_derivatives}
  \begin{aligned}
    \dot{\phi}_0(r) &= \kappa\, \phi'_0(r) - (3\kappa' - \mu)\, \phi_0(r) -
    \alpha_2\, \mu\, \phi_1(r),\\
    \dot{\phi}_1(r) &= \frac{1}{2}\, \alpha_2\, \mu\, \phi_0(r) - \frac{1}{2}\,
    \alpha_0\, \mu\,  \phi_2(r) - \mu\, \phi_1,\\
    \dot{\phi}_2(r) &= \frac{1}{2}\, \alpha_0\, \mu\, \phi_1(r) - \frac{1}{2}\,
    \alpha_0\, \mu\,  \phi_3(r),\\
    \dot{\phi}_3(r) &= \frac{1}{2}\, \alpha_0\, \mu\, \phi_2(r) - \frac{1}{2}\,
    \alpha_2\, \mu\,  \phi_4(r) + \mu\, \phi_3(r),\\
    \dot{\phi}_4(r) &= -\kappa\, \phi'_4(r) + (3\kappa' - \mu)\,\phi_4(r) +
    \alpha_2\, \mu\, \phi_3(r),
  \end{aligned}
\end{equation}
where the values of the fields on the r.h.s can be evaluated from \eqref{constraints},
\eqref{algebr_ic_phi_2}.

\begin{figure}[hbt]
  \centering
  \subfigure[]{
    \includegraphics[height=4.5cm,width=5.5cm]{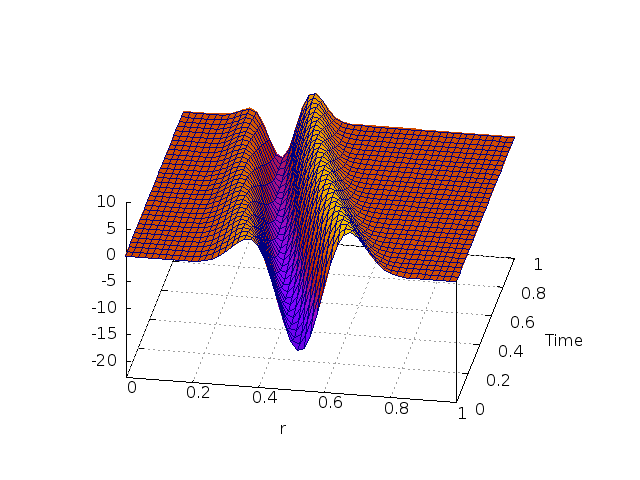}
  }
  \subfigure[]{
    \includegraphics[height=4.5cm,width=5.5cm]{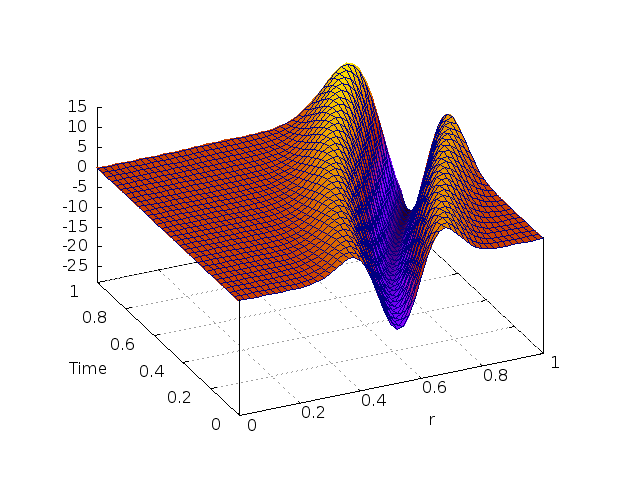}
  }
  \caption{The numerical solutions of (a) $\phi_0$ and (b) $\phi_4$ for the evolution of 
  general initial data---centered around $r=0.5$---for the $l=2$ mode in the `diagonal' 
  representation.}
\label{fig:numer_solut_plots}
\end{figure}

Finally, we explain how we specify the boundary conditions. As discussed in
sec.~\ref{sec:relating_first_second_order_system} we need to use the available information
from the first-order system~\eqref{spin-2_eq}. In this system there is only one component,
namely~$\phi_0$, which propagates inward from the boundary at $r=1$. Thus, there is only
one free function to be specified on that boundary, which characterizes the solution
inside given initial data. This must be the case also for the second order system after
imposing the boundary conditions.

The second order wave equations require for each component a boundary condition at
$r=1$ (recall that $r=0$ is a total characteristic so we cannot prescribe any conditions
there). We impose these conditions in the form of a Robin
condition~\eqref{Robin_boundary}
\begin{equation*}
\phi_i(t,1) + \phi'_i(t,1) = g_i(t), \quad i=0,1,2,3,4.
\end{equation*}
The boundary functions $g_i(t)$ are computed from the first-order system in terms of the
fields and their time derivatives. There is no unique way to do this because there are
both constraint and evolution equations involving $\phi_1$, $\phi_2$ and $\phi_3$ which
could be use for this purpose. We
choose to use the constraint equations~(\ref{eq:1st_c1}-\ref{eq:1st_c3}) for $\phi_1$,
$\phi_2$ and $\phi_3$ and the evolution equations~\eqref{eq:1st_ev0}
and~\eqref{eq:1st_ev4} for $\phi_0$ and $\phi_4$.

Since $\phi_0$ is freely specifiable on the boundary we choose it simply as
zero, so that $\phi_0$ and its time derivative vanish on the boundary. Wherever $\phi_0$
appears in the equations used, we simply drop it. The functions $g_i(t)$ are computed by
evaluating the left hand side of~\eqref{Robin_boundary} with the spatial derivative
substituted from the corresponding first-order equation and putting $\phi_0$ and
$\dot\phi_0$ equal to zero. 

The resulting equation for each component is imposed numerically using the SAT method.
Note, that we have essentially only rewritten the first-order equation for the components
in a superficial form of a Robin condition. We could just as well have chosen a Neumann
condition or even a Sommerfeld condition. In all cases the effective boundary condition,
i.e., the penalty term to be added to the discretized wave equation would have been the
same. In fact, we do not see any difference if we implement the same boundary conditions
as Neumann or even as Sommerfeld condition.

\begin{figure}[hbt]
  \centering
  \subfigure[]{
    \includegraphics[height=4.cm,width=5.5cm]{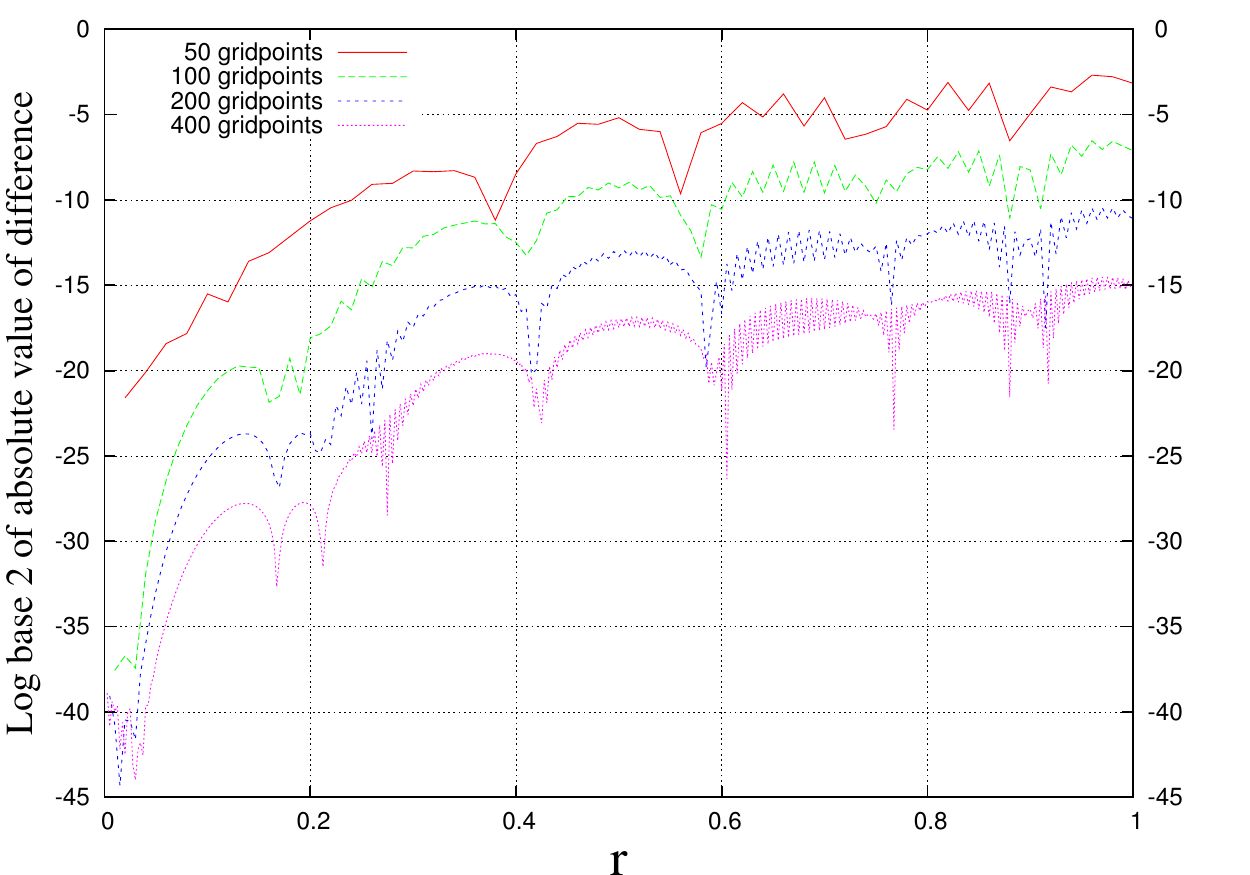}
    \label{subfig:1st_order}}
  \subfigure[]{
    \includegraphics[height=4.cm,width=5.5cm]{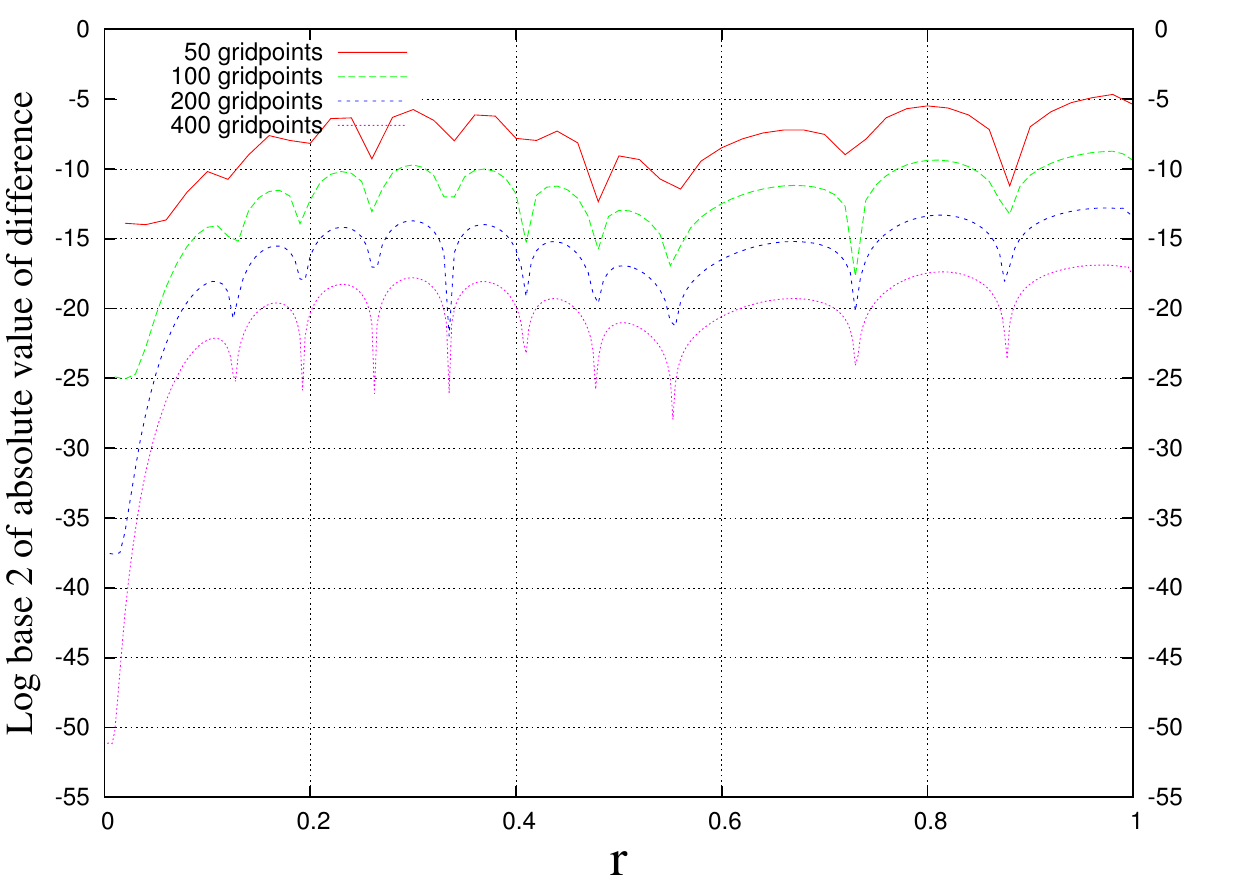}
    \label{subfig:2nd_order}}
  \caption{The convergence plots of $\phi_4$ for the evolution of the initial data \eqref{algebr_ic_phi_2}, 
  \eqref{algebr_ic_derivatives} in the `diagonal' representation at time $t=1$ in (a) the first 
  order approach \cite{Beyer:2012ti} and (b) the second order approach developed in the
  present work.}
\label{fig:conv_comparison_plots}
\end{figure}

We specify initial data for the simplest non-trivial case $l=2$ in the `diagonal'
representation. For the evolution of the data \eqref{algebr_ic_phi_2},
\eqref{algebr_ic_derivatives} described above, the numerical solutions for the `ingoing'
and `outgoing' components $\phi_0$ and $\phi_4$, respectively, are presented in
Fig.~\ref{fig:numer_solut_plots}. These components satisfy first-order advection equations
which are purely ingoing resp. outgoing. These properties are clearly visible in the
plots.

In Fig.~\ref{fig:conv_comparison_plots} we compare the convergence plots of the $\phi_4$
component at $t=1$ obtained from the evolution of the same initial data
\eqref{algebr_ic_phi_2} and \eqref{algebr_ic_derivatives} in the two numerical approaches
developed in the present work and \cite{Beyer:2012ti}. More precisely, in
Fig.~\ref{subfig:1st_order} the initial data are evolved using the first-order system \eqref{eq:1st_ev0}-\eqref{eq:1st_ev4},
while in Fig.~\ref{subfig:2nd_order} the second order system \eqref{spin-2_wave_final} is
used. In each case, the difference to a high resolution run is computed, which here is 800
grid points. In both cases we find 4th order convergence. Notice, though, that in the
second order case the plots look much cleaner. The high frequency features, appearing on
the convergence plot of the first order system, are not present in the plot for the second
order system. This result confirms the prediction of \cite{Kreiss:2002du} that the
spurious waves will disappear in the second-order formulation.

\begin{table}[htb]
  \centering
  \begin{tabular}{|c|cc|cc|}
    \hline
    &  \multicolumn{2}{c|}{$\phi_0$}   & \multicolumn{2}{c|}{$\phi_4$} \\
    \hline
    Grid  & $\log_2(||E||_2)$  &     Rate     & $\log_2(||E||_2)$ &   Rate     \\
    \hline
    50    &     0.4593         &              &     0.5439        &            \\
    \hline 
    100   &    -3.4714         &    3.9307    &    -3.5608        &  4.1047    \\ 
    \hline  
    200   &    -7.4508         &    3.9794    &    -7.4986        &  3.9378    \\ 
    \hline
    400   &    -11.4352        &    3.9844    &    -11.4859       &  3.9873    \\
    \hline
  \end{tabular}
  \caption{Convergence of the solutions to the wave equations towards a solution at high
    resolution (800 grid points) of the first order equation. The table shows the
    (logarithms of) the normalized $l^2$ norm of the absolute differences and the corresponding
    convergence rates at time $t=1$.}
\label{tab:conv_rate_comp}
\end{table}
Now, we reproduce numerically the analytic result of
sec.~\ref{sec:relating_first_second_order_system}: we show that if the initial data
satisfy the constraints \eqref{constraints} of the first order system and are evolved with
the second order system \eqref{spin-2_wave_final}, then the resulting numerical solutions
converge (with increasing resolution) to the one obtained by evolving the same
initial data with the evolution equations \eqref{eq:1st_ev0}-\eqref{eq:1st_ev4} 
of the first order system. To make a connection
with the results of Fig.~\ref{fig:conv_comparison_plots}, we consider again initial data
centered around $r=0.5$ of the form \eqref{algebr_ic_phi_2}, \eqref{algebr_ic_derivatives}
and impose boundary conditions as described above. We evolve these data
with both evolution systems and compute the absolute error between the
numerical solutions of the second order approach and the numerical solution of the highest
resolution (800 grid points) in the first order approach. The normalized $l^2$ norm of the
computed absolute error and the corresponding convergence rates at time $t=1$ for the
components $\phi_0, \phi_4$ are listed in Tab.~\ref{tab:conv_rate_comp}. We find that the
solutions agree within numerical accuracy. This confirms the statement made in
sec.~\ref{sec:relating_first_second_order_system} that the two systems have identical
solutions given the same initial data.

\section{Discussion}
\label{sec:disscusion}

In this work we have reformulated the first-order system for the spin-2 field equations
for linear gravitational perturbations on Minkowski space as a system of coupled
second-order wave equations. The system was implemented numerically and studied
under certain simplifying assumptions.

Our first analytical result, see sec.~\ref{sec:relating_first_second_order_system}, 
is that the first-order system and the second-order system are
equivalent if and only if the underlying space-time is conformally flat and has a
vanishing scalar curvature. In the case under study the space-time is conformal to
Minkowski space-time, i.e., conformally flat and the conformal factor has been chosen in 
such a way that the scalar curvature vanishes. Thus, we can evolve the spin-2 field using
either the first or the second order system. Given initial and boundary data computed from
the first-order system the second-order system provides the same solutions.

Our main goal was the comparison of the numerical properties of the two formulations in
the conformal setting given in sec.~\ref{sec:minkowski_near_i}. From the numerical results
presented in sections~\ref{sec:exact_solution} and \ref{sec:general_data}  we can conclude
that we get the same qualitative behaviour of the numerical solutions. In particular, we
have shown explicitly that solutions of the second-order system converge to a solution of
the first-order system with the same initial and boundary data. The solutions can be
evolved stably up to and including $I^+$ in the diagonal case but not beyond. Any attempt
to evolve beyond $I^+$ with the same algorithm end in numerical instability. In a recent 
paper \cite{Beyer:2013} we showed that in the first order formulation of the spin-2 zero-rest-mass 
equation developed in \cite{Beyer:2012ti} it is possible to evolve asymptotically Euclidean 
initial data beyond $I^+$ and extract the physically important radiation fields on $\scri^+$.

Concerning the comparison of the two approaches we confirm the claims
in~\cite{Kreiss:2002du}: with the same number of grid points we obtain better accuracy
roughly by a factor 4 with the second-order system and we do not see any spurious
high-frequency waves in the second-order system.

In the present work we carried out numerical simulations using (first and second order) SBP operators of the same (full or diagonal) norm. Our results agree with their designed accuracy as described in~\cite{Mattsson:2004br}. In addition, in order to explore the potentialities of the available SBP operators, we went one step further and combined first and second order discrete SBP operators with respect to \emph{different} inner products. Our numerical results indicate that the requirement of using SBP operators with the same norm does not seem to be necessary: the second order full norm operator seems to perform quite well even when combined with first order SBP operators based on a restricted full norm.

\section{Acknowledgments}
\label{sec:acknowledgments}

This work has been supported by Marsden grant UOO0922 of the Royal Society of New Zealand. The authors are grateful to H. Friedrich, O. Reula and M. Tiglio for discussions of separate aspects of this research.


\appendix

\section{The first order equations}
\label{sec:first-order-equat}

As a reference, we list here the spin-2 zero-rest-mass equation~\eqref{spin-2_eq} in the formulation used
in~\cite{Beyer:2012ti}. The geometry and the notation is exactly the same as described in
sec.~\ref{sec:coord-repr-second}. The system consists of eight equations which can be
split into the five time evolution equations 
\begin{align}
(1+t\kappa') \del_t \phi_0 - \kappa \del_r \phi_0 &= -(3\kappa' - \mu) \phi_0   -\mu \alpha_2 \phi_1 ,\label{eq:1st_ev0}\\
\del_t \phi_1  &=  -  \mu \phi_1 +  \frac12 \mu \alpha_2 \phi_0 - \frac12 \mu \alpha_0 \phi_2,\label{eq:1st_ev1}\\
\del_t \phi_2  &=    \frac12 \mu \alpha_0 \phi_1 - \frac12 \mu \alpha_0 \phi_3,\label{eq:1st_ev2}\\
\del_t \phi_3  &=   \mu \phi_3 +  \frac12 \mu \alpha_0 \phi_2 - \frac12 \mu \alpha_2 \phi_4,\label{eq:1st_ev3}\\
 (1-t\kappa') \del_t \phi_4 + \kappa \del_r \phi_4 &=  (3\kappa' -  \mu) \phi_4 +  \mu \alpha_2 \phi_3\label{eq:1st_ev4}
\end{align}
and the three constraint equations
\begin{align}
- 2 \kappa \del_r \phi_1 + 6 r \mu' \phi_1 - 2 t \kappa' \mu \phi_1 +  \alpha_0 \mu (1-t\kappa') \phi_2 + \alpha_2 \mu (1+t\kappa') \phi_0 &= 0,\label{eq:1st_c1}\\
  -2 \kappa \del_r \phi_2 + 6 r \mu' \phi_2  + \alpha_0 \mu (1-t\kappa') \phi_3 + \alpha_0 \mu (1+t\kappa') \phi_1 &= 0,\label{eq:1st_c2}\\
  - 2 \kappa \del_r \phi_3 +  6 r \mu' \phi_3 + 2 t \kappa' \mu \phi_3 + \alpha_0 \mu (1+t\kappa') \phi_2 + \alpha_2 \mu (1-t\kappa') \phi_4 &= 0.\label{eq:1st_c3}
\end{align}


\end{document}